\documentclass[journal]{IEEEtran}
\usepackage{graphicx}     
\usepackage{amsmath}       
\usepackage{booktabs}       
\usepackage{caption}       
\usepackage{biblatex}      
\usepackage{algorithm}      
\usepackage{orcidlink}
\usepackage{algorithmic}   
\usepackage{adjustbox}   
\usepackage{amssymb}
\usepackage{array}          
\usepackage{float}    
\usepackage[caption=false,font=footnotesize]{subfig}
\usepackage{verbatim} 
\usepackage[compact]{titlesec}
\titlespacing*{\section}{0pt}{*0}{*0}

\title{EAQKD: Entanglement-Based Authenticated Quantum Key Distribution}

\author{
    \IEEEauthorblockN{
        Noureldin Mohamed\IEEEauthorrefmark{1}\orcidlink{0009-0001-4150-8690}\thanks{Corresponding author: nomo89098@hbku.edu.qa} (nomo89098@hbku.edu.qa),
        Saif Al-Kuwari\IEEEauthorrefmark{1}\orcidlink{0000-0002-4402-7710} (smalkuwari@hbku.edu.qa)
    }
    \\
    \IEEEauthorblockA{
        \IEEEauthorrefmark{1}Qatar Center for Quantum Computing, College of Science and Engineering, Hamad Bin Khalifa University, Doha, Qatar.
    }
}

\date{}

\begin{document}

\maketitle

\begin{abstract}
The promise of unconditional security in the Quantum Key Distribution (QKD) depends on the availability of an authenticated classical channel. However, practical implementations often overlook this requirement or rely on computational assumptions that compromise long-term security.  To overcome these challenges, this paper presents Entanglement-Based Authenticated Quantum Key Distribution (EAQKD), a novel protocol that addresses critical security and practical limitations in quantum cryptographic key exchange. Our approach integrates quantum entanglement distribution with information-theoretic authentication. We evaluate EAQKD's performance through a comprehensive discrete-event simulation framework modeled on realistic channel characteristics and experimental device parameters. Our modeling incorporates parameters from practical quantum optics setups, including SPDC entanglement sources, superconducting nanowire detectors, and fiber channel imperfections. Our results show quantum bit error rates consistently below the 11\% security threshold (ranging from 1.86\% at 10 km to 9.27\% at 200 km), with secure key rates achieving $1.12 \times 10^5$ bits/s at short distances and maintaining practical rates of 9.8 bits/s at 200 km.
When integrated with quantum repeater architectures, our analysis projects that EAQKD can extend secure communication beyond 500 km while providing information-theoretic security guarantees. Comparative analysis against the BB84, E91, and Twin-Field QKD protocols demonstrates EAQKD's superior balance of security, practical performance, and implementation robustness. This work advances quantum cryptography by providing a rigorously analyzed engineering reference for secure key distribution in future quantum communication networks.
\end{abstract}


\section{Introduction}
Quantum Key Distribution (QKD) represents a revolutionary approach to cryptographic key exchange, harnessing the intrinsic properties of quantum mechanics to achieve information-theoretic security \cite{bennett2014quantum}. Unlike classical cryptographic methods, such as RSA and Diffie-Hellman, QKD's security is grounded in fundamental physical laws rather than computational complexity assumptions. This distinction becomes increasingly important as advances in quantum computing threaten to compromise traditional cryptographic systems \cite{shor2000simple}.

Despite significant theoretical advances in QKD, practical implementations face several critical challenges that limit their security and deployability \cite{scarani2009security}. Prepare-and-measure protocols like BB84 \cite{bennett1984quantum} remain vulnerable to photon-number-splitting attacks when implemented with attenuated laser sources \cite{hwang2003quantum}. Additionally, detector-based side-channel attacks have been demonstrated against multiple QKD systems, compromising their security guarantees \cite{lydersen2010hacking}. Similarly, the classical communication channel used for post-processing is often inadequately authenticated in practice, creating a critical vulnerability where the authentication assumption does not match the implementation~\cite{portmann2022security}. Without robust authentication, an adversary can impersonate both parties, effectively establishing separate quantum channels with Alice and Bob while controlling all classical communication between them (a Man-in-the-Middle attack). The security of this authentication is strictly bound by the tag length $t$ and the probability of successful forgery~\cite{krawczyk2002hmac}:
\begin{equation}
P_{\text{forge}} \leq \frac{l \cdot q}{2^t}
\end{equation}
where $l$ is the message length and $q$ is the number of authentication queries. Finally, transmission losses in optical fibers limit the practical range of most QKD systems to approximately 100-200 km without quantum repeaters~\cite{takeoka2014fundamental}.
While entanglement-based protocols like E91 \cite{ekert1991quantum} address some of these challenges through Bell inequality verification, they typically achieve lower key rates than prepare-and-measure approaches and still require proper authentication of the classical channel \cite{ma2007quantum}. Recent advances, such as Twin-Field QKD \cite{lucamarini2018overcoming}, improve the distance scaling but introduce new implementation challenges related to phase stabilization and interferometric alignment.

Despite these significant theoretical advances, a notable gap remains in addressing both quantum and classical vulnerabilities simultaneously while maintaining practical key rates over useful distances. Most current approaches either prioritize quantum security at the expense of rate (e.g., standard entanglement protocols) or focus on distance at the expense of implementation complexity (e.g., TF-QKD), often leaving the classical authentication layer relying on computational assumptions. Our research addresses this gap by developing the EAQKD protocol, which integrates entanglement-based quantum security with robust information-theoretic classical authentication while optimizing for practical performance.

In this paper, we introduce Entanglement-Based Authenticated Quantum Key Distribution (EAQKD), a novel protocol that addresses these limitations by combining the security advantages of entanglement-based QKD with robust classical authentication. Our key contributions include:

\begin{itemize}
    \item A comprehensive protocol design that integrates quantum entanglement distribution with quantum-resistant authentication of the classical channel
    \item Asymmetric basis selection optimization that significantly improves key generation efficiency while maintaining security verification
    \item Comprehensive simulation-based evaluation of the protocol's performance across distances from 10 km to 200 km, with detailed analysis of quantum bit error rates, Bell parameter measurements, and secure key rates.
    \item Implementation of entanglement purification techniques that extend the protocol's effective range
    \item Comparative evaluation against leading QKD protocols, demonstrating EAQKD's advantages in security, distance, and practical deployability
    \item Analysis of EAQKD's performance when enhanced with quantum repeaters, showing potential for secure key distribution beyond 500 km.
\end{itemize}

Our simulation results demonstrate that EAQKD maintains quantum bit error rates (QBER) below the critical security threshold of 11\% across all tested distances, with values ranging from 1.86\% at 10 km to 9.27\% at 200 km. Bell inequality measurements confirm genuine quantum entanglement with $S = 2.61 \pm 0.08$ at 10 km, decreasing to $S = 2.12 \pm 0.11$ at 150 km. The secure key rate achieves $1.12 \times 10^5$ bits/second at short distances and maintains practical rates of approximately 10 bits/second at 200 km.

The remainder of this paper is organized as follows: Section \ref{sec:preliminaries} provides essential background on QKD and the limitations of existing approaches. Section \ref{sec:protocol} details the EAQKD protocol design, outlining the five operational phases from entanglement generation to authentication key renewal. Section \ref{sec:security_simulation} presents our consolidated security analysis and the discrete-event simulation framework used to validate the system. Section \ref{sec:performance} analyzes the performance results, focusing on key rates, entanglement quality, and purification effects. Section \ref{sec:benchmarking} offers a comparative evaluation against the BB84, E91, and Twin-Field QKD protocols. Section \ref{sec:repeaters} examines the feasibility of range extension using quantum repeaters, and Section \ref{sec:conclusion} concludes with a summary of our findings and directions for future work.

\section{Preliminaries}
\label{sec:preliminaries}
Quantum Key Distribution (QKD) has evolved significantly since its inception, with each protocol generation addressing specific vulnerabilities while introducing new capabilities \cite{pirandola2020advances}. The security of QKD derives from fundamental quantum principles, particularly the measurement disturbance principle formalized as:

\begin{equation}
\Delta X \cdot \Delta P \geq \frac{\hbar}{2}
\end{equation}

where $\Delta X$ and $\Delta P$ represent uncertainties in complementary observables. This principle ensures that any measurement attempt by an eavesdropper necessarily introduces detectable disturbances \cite{heisenberg1927uncertainty}.

\subsection{Prepare-and-Measure QKD Protocols}

The seminal BB84 protocol \cite{bennett1984quantum} established the prepare-and-measure paradigm for QKD. In BB84, Alice encodes classical bits into the polarization states of single photons, selecting randomly between two conjugate bases: the rectilinear basis $\{|0\rangle, |1\rangle\}$ and the diagonal basis $\{|+\rangle = \frac{|0\rangle+|1\rangle}{\sqrt{2}}, |-\rangle = \frac{|0\rangle-|1\rangle}{\sqrt{2}}\}$. Bob measures each received photon in a basis chosen independently and randomly. After transmission, Alice and Bob publicly disclose their basis choices over an authenticated classical channel, retaining only the bits where their bases align to form a sifted key.

The security of BB84 relies on the fact that measuring a quantum state in the wrong basis introduces errors. The quantum bit error rate (QBER) serves as the primary security parameter, with the secure key rate $R$ given by:
\begin{equation}
R_{\text{BB84}} = R_{\text{raw}} \cdot [1 - 2h(e)] - f(e) \cdot h(e)
\end{equation}
where $R_{\text{raw}} = \eta\mu e^{-\mu}/2$ is the raw key rate, $\eta$ is the channel transmittance, $\mu$ is the mean photon number, $h(e) = -e \log_2(e) - (1-e) \log_2(1-e)$ is the binary Shannon entropy, and $f(e) \geq 1$ is the error correction inefficiency \cite{scarani2009security}.

However, prepare-and-measure protocols are susceptible to several vulnerabilities:

\begin{itemize}
    \item \textbf{Photon-Number-Splitting (PNS) Attacks}: Practical single-photon sources often emit pulses with Poisson-distributed photon numbers. Eve can exploit multi-photon pulses by capturing excess photons without detection \cite{hwang2003quantum}. For a source with mean photon number $\mu$, the probability of emitting multiple photons is:
    \begin{equation}
    P(n \geq 2|\text{click}) = \frac{1 - e^{-\mu} - \mu e^{-\mu}}{1 - e^{-\mu}}
    \end{equation}
    \item \textbf{Detector Vulnerabilities}: Time-shift attacks and blinding attacks exploit imperfections in single-photon detectors, allowing Eve to control Bob's measurement outcomes without detection \cite{lydersen2010hacking}.
    \item \textbf{Distance Limitations}: Photon loss in optical fibers follows an exponential decay:
    \begin{equation}
    \eta_{\text{channel}} = 10^{-\alpha d/10}
    \end{equation}
    where $\alpha$ is the attenuation coefficient (typically 0.2 dB/km) and $d$ is the distance in kilometers. This fundamental limitation restricts BB84's practical range to approximately 100-200 km without quantum repeaters \cite{takeoka2014fundamental}.
\end{itemize}

Decoy-state protocols \cite{hwang2003quantum} partially address the PNS vulnerability by varying the intensity of transmitted pulses, allowing Alice and Bob to detect photon-number-dependent attacks. However, they do not address detector vulnerabilities or the authenticated channel requirement.

\subsection{Entanglement-Based QKD Protocols}
Original entanglement-based protocols like E91 leverage quantum correlations through Bell states \cite{ekert1991quantum}. Our implementation follows the BBM92 variant of the entanglement scheme \cite{bennett1984quantum}, which simplifies the measurement strategy to two conjugate bases ($\sigma_z, \sigma_x$) while retaining the security properties derived from Bell inequality violations as a statistical diagnostic.

For the singlet state:
\begin{equation}
|\psi^-\rangle = \frac{1}{\sqrt{2}}(|0\rangle_A|1\rangle_B - |1\rangle_A|0\rangle_B)
\end{equation}
the correlation function for measurements along directions $\vec{a}$ and $\vec{b}$ is:
\begin{equation}
E(\vec{a},\vec{b}) = -\vec{a} \cdot \vec{b} = -\cos(\theta_{ab})
\end{equation}
where $\theta_{ab}$ is the angle between measurement directions. Security is verified through Bell inequality violations, with the CHSH parameter:
\begin{equation}
S = |E(\vec{a}_1,\vec{b}_1) + E(\vec{a}_1,\vec{b}_2) + E(\vec{a}_2,\vec{b}_1) - E(\vec{a}_2,\vec{b}_2)|
\end{equation}
exceeding the classical limit of 2 for genuine quantum entanglement. The theoretical maximum for quantum systems is $S = 2\sqrt{2} \approx 2.83$ \cite{clauser1969proposed}.

The secure key rate is derived from the estimated phase errors. However, for diagnostic purposes under a symmetric noise model (e.g., Werner states), the quantum bit error rate $e$ relates to the Bell parameter $S$ by:
\begin{equation}
e = \frac{1}{2} - \frac{S}{4\sqrt{2}}
\end{equation}
While entanglement-based protocols address source-side loopholes, they face challenges in entanglement source complexity and detector efficiency requirements \cite{ma2007quantum}.

\subsection{Twin-Field QKD}
The fundamental rate-distance limitation in QKD is characterized by the repeaterless PLOB bound \cite{pirandola2017fundamental}:
\begin{equation}
R \leq -\log_2(1-\eta)
\end{equation}
which for small $\eta$ approximates to $R \lesssim 1.44\eta$ bits per channel use.

Twin-Field QKD (TF-QKD) uses single-photon interference at a middle station to overcome this limit, achieving scaling proportional to $\sqrt{\eta}$ rather than $\eta$ \cite{lucamarini2018overcoming}:
\begin{equation}
R_{\text{TF}} \propto \sqrt{\eta_A \eta_B} \approx \eta_{\text{total}}^{1/2}
\end{equation}
This allows for longer distances but introduces significant implementation complexity regarding phase stabilization and potential side channels related to phase reference leakage \cite{minder2019experimental}.

\section{EAQKD Protocol}
\label{sec:protocol}
In this paper, we introduce Entanglement-Based Authenticated Quantum Key Distribution (EAQKD), which integrates entanglement-based quantum key distribution with quantum-resistant authentication to address both quantum and classical vulnerabilities while maintaining practical performance. The protocol consists of five main phases: entanglement generation, quantum distribution, measurement, classical post-processing, and authentication. Algorithm \ref{alg:eaqkd} provides a pseudocode for EAQKD. 

\begin{algorithm}
\caption{EAQKD Protocol}
\label{alg:eaqkd}
\begin{algorithmic}[1]
\REQUIRE Security parameter $\lambda$, Fiber length $L$, Bit error threshold $\epsilon_{\text{bit}}$, Phase error threshold $\epsilon_{\text{phase}}$
\ENSURE Secret key $K_{\text{sec}}$

\STATE \textbf{Entanglement Generation:}
\STATE Generate Bell state $|\psi^-\rangle = \frac{1}{\sqrt{2}}(|01\rangle - |10\rangle)$
\STATE Apply phase compensation using $U_{\phi} = e^{i\hat{H}_{\text{comp}}t}$
\STATE Verify fidelity $F \geq 1 - 2^{-\lambda/2}$
\STATE Apply purification if needed (based on distance/fidelity)

\STATE \textbf{Quantum Distribution:}
\STATE Calculate channel loss $\eta = 10^{-\alpha L/10}$
\STATE Distribute entangled photons to Alice and Bob
\STATE Implement active polarization tracking

\STATE \textbf{Measurement Phase:}
\STATE Set basis probabilities $p_z = 0.9$, $p_x = 0.1$
\STATE Alice and Bob independently choose measurement bases
\FOR{$i = 1$ to $N$}
    \STATE Alice measures in basis $\mathbb{B}_A^i \in \{\sigma_z, \sigma_x\}$
    \STATE Bob measures in basis $\mathbb{B}_B^i \in \{\sigma_z, \sigma_x\}$
    \STATE Record outcomes $m_A^i$ and $m_B^i$
\ENDFOR

\STATE \textbf{Basis Reconciliation (Authenticated):}
\STATE Alice and Bob announce basis choices (Authenticated via Wegman-Carter)
\STATE Identify matching $\sigma_z$ bases: $\mathcal{M} = \{i : \mathbb{B}_A^i = \mathbb{B}_B^i = \sigma_z\}$
\STATE Extract raw keys: $K_A^{\text{raw}} = \{m_A^i : i \in \mathcal{M}\}$, $K_B^{\text{raw}} = \{m_B^i : i \in \mathcal{M}\}$
\STATE Identify test bits: $\mathcal{T} = \{i : \mathbb{B}_A^i = \mathbb{B}_B^i = \sigma_x\}$

\STATE \textbf{Parameter Estimation:}
\STATE Calculate bit error rate: $e_z = \frac{1}{|\mathcal{M}|}\sum_{i \in \mathcal{M}} \delta(m_A^i, m_B^i \oplus 1)$
\STATE Calculate phase error rate: $e_x = \frac{1}{|\mathcal{T}|}\sum_{i \in \mathcal{T}} \delta(m_A^i, m_B^i \oplus 1)$
\STATE Infer Bell parameter (Diagnostic): $S \approx 2\sqrt{2}(1-2e_x)$
\IF{$e_z > \epsilon_{\text{bit}}$ OR $e_x > \epsilon_{\text{phase}}$}
    \STATE \textbf{Abort protocol}
\ENDIF

\STATE \textbf{Error Correction:}
\STATE Select LDPC code rate $R$ based on $e_z$
\STATE Alice computes syndrome $s_A = H \cdot K_A^{\text{raw}}$
\STATE Alice sends authenticated syndrome to Bob
\STATE Bob performs decoding: $K_B^{\text{cor}} = \text{LDPC-Decode}(K_B^{\text{raw}}, s_A, H)$
\STATE Calculate information leakage $\text{leak}_{\text{EC}}$

\STATE \textbf{Privacy Amplification:}
\STATE Calculate finite-size penalty $\Delta_{\text{fs}}$
\STATE Calculate secure key length $\ell = |K_B^{\text{cor}}| \cdot [1 - h(e_x)] - \text{leak}_{\text{EC}} - \Delta_{\text{fs}}$
\STATE Apply Toeplitz hash function: $K_{\text{sec}} = T \cdot K_B^{\text{cor}}$

\STATE \textbf{Authentication Key Update:}
\STATE Reserve portion for next round authentication: $K_{\text{auth}}^{\text{new}} = \text{First}(K_{\text{sec}}, 4\lambda)$
\STATE Output Secret Key: $K = \text{Rest}(K_{\text{sec}})$
\STATE Update Authentication Key: $K_{\text{auth}} \leftarrow K_{\text{auth}}^{\text{new}}$ (for Wegman-Carter OTP)

\RETURN $K$
\end{algorithmic}
\end{algorithm}

\begin{figure*}[ht]
\centering
\includegraphics[width=0.9\linewidth]{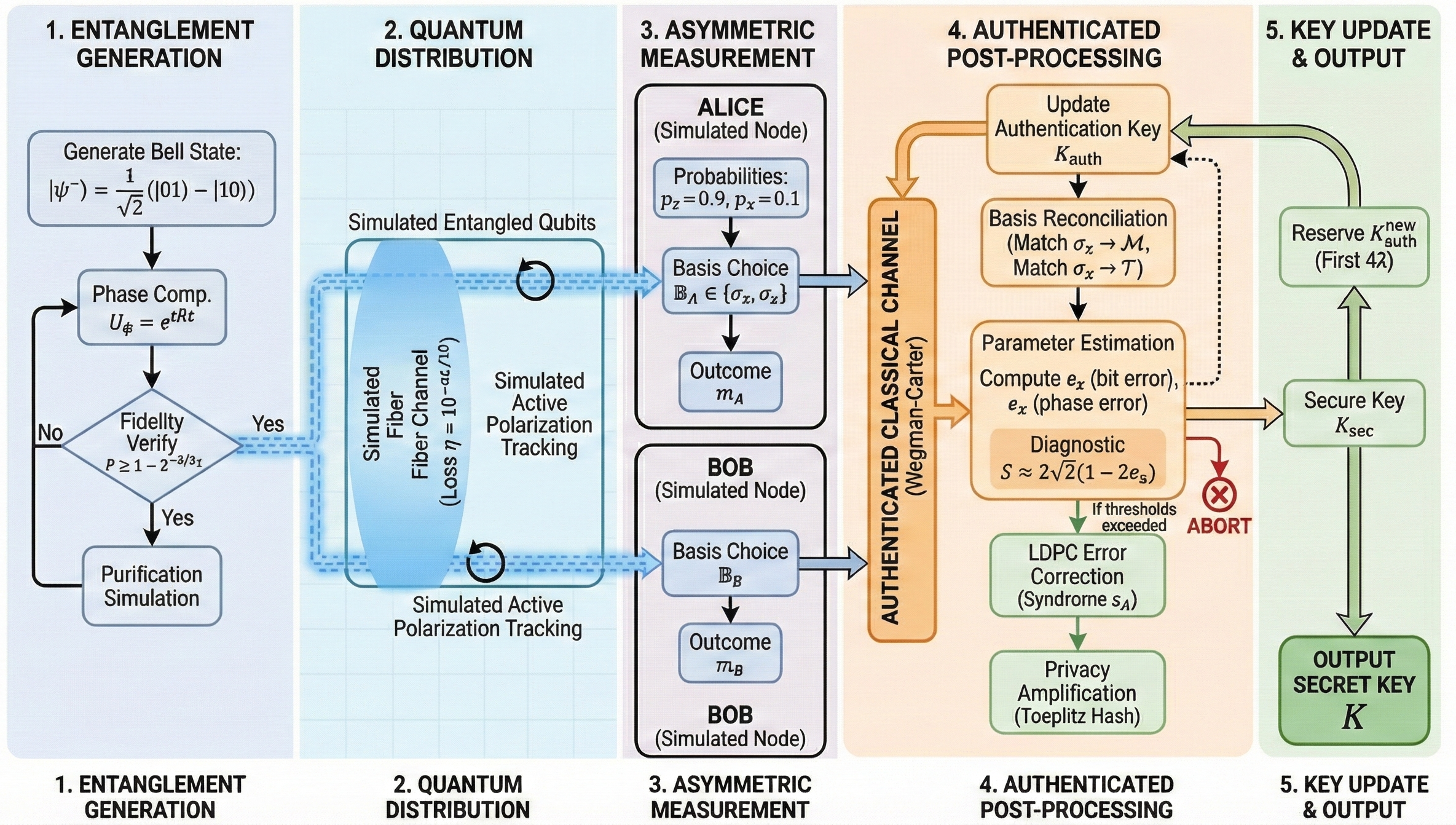} 
\caption{Comprehensive workflow of the EAQKD protocol. The diagram highlights the five operational phases: (1) Entanglement Generation and optional purification, (2) Quantum Distribution with active compensation, (3) Asymmetric Measurement ($p_z=0.9$), (4) Authenticated Classical Post-Processing using Wegman-Carter authentication and strictly separate error checks, and (5) Key Update/Output. The Bell parameter $S$ is utilized solely as a source diagnostic tool.}
\label{fig:eaqkd_workflow}
\end{figure*}

To visualize the architectural flow and the interaction between quantum and classical layers, Figure \ref{fig:eaqkd_workflow} depicts the end-to-end operation of the EAQKD protocol. The process is segmented into five distinct phases, moving from the physical generation of purified Bell pairs (Phase 1) and their distribution over fiber channels with active polarization compensation (Phase 2), to the asymmetric measurement strategy (Phase 3). Crucially, Phase 4 illustrates the rigorous post-processing pipeline where basis reconciliation is secured via Wegman-Carter authentication, and security is enforced by strictly checking phase errors ($e_x$) against the finite-size security threshold. Finally, Phase 5 executes the automated renewal of the authentication key, ensuring composable security for subsequent rounds.

\subsection{Phase 1: Entanglement Generation}
The protocol begins with the generation of high-fidelity Bell states $|\psi^-\rangle = \frac{1}{\sqrt{2}}(|01\rangle - |10\rangle)$. We selected this state for its robustness against collective noise and its perfect anti-correlation properties in complementary measurement bases. We model the entanglement source as a Spontaneous Parametric Down-Conversion (SPDC) system using a periodically poled potassium titanyl phosphate (PPKTP) crystal pumped by a 405 nm laser. The simulated source produces entangled photon pairs at telecom wavelength (1550 nm) with a generation rate of approximately $10^6$ pairs per second and a fidelity of 0.98 to the target Bell state.

For distances exceeding 100 km, we implement the DEJMPS entanglement purification protocol to improve the fidelity of the distributed states. This process consumes additional entangled pairs to produce fewer pairs with higher fidelity, trading off generation rate for improved state quality. The purification process is applied when the estimated fidelity falls below 0.95, as determined through simulated quantum state tomography on a subset of the generated pairs.

\subsection{Phase 2: Quantum Distribution}
The entangled photons are distributed to Alice and Bob through standard telecom fibers with a modeled attenuation coefficient of $\alpha = 0.2$ dB/km. To counteract channel-induced disturbances, the simulation implements active polarization compensation using liquid crystal retarders controlled by a feedback system that monitors the polarization state of reference pulses sent periodically through the same fiber. This ensures the integrity of the quantum states upon arrival at the measurement stations.

\subsection{Phase 3: Measurement}
Upon receiving their respective photons, Alice and Bob independently select measurement bases according to an asymmetric probability distribution: the $\sigma_z$ (computational) basis with probability $p_z = 0.9$ and the $\sigma_x$ (Hadamard) basis with probability $p_x = 0.1$. This asymmetric selection significantly improves the key generation efficiency compared to the symmetric case ($p_z = p_x = 0.5$) used in the original E91 protocol, while still providing sufficient statistics for security verification through Bell inequality testing. The measurement bases are defined as:
\begin{align}
\sigma_z &: \{|0\rangle, |1\rangle\} \\
\sigma_x &: \left\{|+\rangle = \frac{|0\rangle + |1\rangle}{\sqrt{2}}, |-\rangle = \frac{|0\rangle - |1\rangle}{\sqrt{2}}\right\}
\end{align}
For the singlet state $|\psi^-\rangle$, the joint measurement probabilities exhibit perfect anti-correlation in both bases. Deviations from these ideal probabilities indicate either channel noise or eavesdropping. The detection system is modeled using polarizing beam splitters and superconducting nanowire single-photon detectors (SNSPDs) with a detection efficiency $\eta_d = 0.85$ and a dark count rate $r_d = 10$ Hz.

\subsection{Phase 4: Classical Post-Processing}
After the quantum phase is complete, Alice and Bob proceed with classical post-processing over an authenticated public channel. This phase focuses on extracting a secure key from the raw measurement data.

First, they identify the subset of measurements where they both used the $\sigma_z$ basis to form their raw key, and the subset where they both used the $\sigma_x$ basis for security verification. Given the asymmetric basis selection probabilities ($p_z = 0.9$), approximately 81\% of the successful coincidence detections contribute to the sifted key. The quantum bit error rate (QBER) is estimated in the $\sigma_z$ basis ($e_z$) for error correction and in the $\sigma_x$ basis ($e_x$) for privacy amplification. For system diagnostic purposes, we infer the Bell parameter $S$ from the test basis error rate under the assumption of a rotationally symmetric Werner state:
\begin{equation}
S_{\text{diag}} = 2\sqrt{2}(1 - 2e_x)
\end{equation}
If $e_z$ or $e_x$ exceed their respective security thresholds, the protocol aborts.

For error correction, we implement rate-adaptive low-density parity-check (LDPC) codes. In our simulations, the error correction efficiency $f(e_z)$ is observed to be approximately $1.05-1.1$. The total information leakage during reconciliation is $\text{leak}_{\text{EC}} = |s_A| + |t_{\text{ver}}|$, corresponding to the syndrome length and verification tag. Finally, privacy amplification is applied using Toeplitz matrix hashing to eliminate residual information. The final secure key length $\ell$ is calculated as:
\begin{equation}
\ell = n_z [1 - h(e_z) - h(e_x)] - \text{leak}_{\text{EC}} - \Delta_{\text{fs}} - \log_2\frac{1}{\epsilon_{\text{cor}}}
\end{equation}
where $\Delta_{\text{fs}}$ accounts for finite-size statistical fluctuations.

\subsection{Phase 5: Authentication and Key Renewal}
To guarantee composable security compatible with the quantum layer, we implement information-theoretic authentication for all classical communication (specifically the basis reconciliation and syndrome exchange in Phase 4). We utilize a Wegman-Carter authentication scheme. For a message $M$, the authentication tag $t$ is generated using a pre-shared key $K_{\text{auth}} = (k_h, k_{\text{otp}})$:
\begin{equation}
t = h_{k_h}(M) \oplus k_{\text{otp}}
\end{equation}
where $h_{k_h}$ is selected from an $\epsilon$-almost strongly universal ($\epsilon$-ASU) family of hash functions (implemented here using polynomial hashing over $GF(2^{128})$), and $k_{\text{otp}}$ is a one-time pad string used for encryption.

To maintain continuous authentication without manual re-keying, the protocol concludes by reserving a portion of the newly generated secure key to refresh the authentication key for the subsequent session:
\begin{equation}
K_{\text{auth}}^{\text{next}} = \text{Extract}(K_{\text{sec}}, L_{\text{auth}})
\end{equation}
This renewal replenishes the one-time pad $k_{\text{otp}}$ required for the Wegman-Carter scheme, ensuring that the authentication remains information-theoretically secure for all future rounds.

\section{Simulation Framework}
\label{sec:security_simulation}
We present a comprehensive framework that integrates our composable security analysis with the discrete-event simulation methodology used to validate EAQKD. This section details the theoretical security definitions, the logical protocol stack, and the physical device parameters modeled within the simulation environment.

\subsection{Security Model}
We adopt a standard security notion relying on the assumptions that an initially authenticated classical channel exists for key exchange, Alice and Bob's laboratories are trusted (with simulated devices characterized to operate without unmodeled side channels), and the adversary is limited only by the laws of quantum mechanics.

The total failure probability of the protocol is bounded by a global security parameter $\varepsilon_{\text{tot}} \approx 2^{-\lambda}$. We select $\lambda \ge 40$ to ensure $\varepsilon_{\text{tot}} \lesssim 10^{-12}$. This composable bound aggregates the failure probabilities across all sub-protocols, ensuring that the final key is secure even when these components are composed together:
\begin{equation}
\varepsilon_{\text{tot}} = \varepsilon_{\text{pe}} + \varepsilon_{\text{EC}} + \varepsilon_{\text{auth}} + \varepsilon_{\text{sec}} + \varepsilon_{\text{cor}}
\end{equation}
The asymptotic key rate under collective attacks is derived via the equivalence to entanglement distillation~\cite{bennett1996purification}. According to the Devetak--Winter bound, the rate is $r \ge 1 - h(e_{\text{bit}}) - h(e_{\text{ph}})$, where $h(\cdot)$ is the binary entropy function, $e_{\text{bit}}$ is the bit error rate observed directly from the simulated data, and $e_{\text{ph}}$ is the phase error rate inferred from the complementary basis statistics.

To account for realistic implementation constraints, we apply post-selection techniques that elevate this collective-attack security to general attacks with finite-size overhead~\cite{christandl2009postselection}. For a finite block of $n_Z$ sifted bits, the final secure key length $\ell$ is derived as:
\begin{equation}
\ell \ge n_Z[1 - h(e_{\text{bit}}) - h(e_{\text{ph}})] - \text{leak}_{\text{EC}} - \Delta_{\text{fs}}
\end{equation}
Here, $\text{leak}_{\text{EC}}$ represents the information leakage during error correction (syndrome + verification tag), and $\Delta_{\text{fs}} \approx O(\log(1/\varepsilon_{\text{sec}}))$ accounts for the statistical fluctuations and smooth min-entropy estimation penalties required for finite-size security~\cite{tomamichel2012tight}.

\subsection{Protocol Optimization and Implementation}
To maximize the secure key rate within this framework, the protocol logic implements four critical optimizations regarding basis selection, error correction, and authentication.

\paragraph{Asymmetric Basis Selection} We simulate an asymmetric basis selection strategy~\cite{lo2005efficient} where the probability of choosing the Z-basis, $p_Z$, is optimized based on the block size $N$. The optimal probability is calculated as:
\begin{equation}
p_Z = 1 - \sqrt{\frac{c \log(1/\varepsilon_{\text{pe}})}{N}}
\end{equation}
For a typical block size of $N=10^8$, we set $p_Z \approx 0.9$. This significantly increases the sifted key fraction to approximately $81\%$, compared to the $25\%$ limit inherent in symmetric ($p_Z=0.5$) protocols.

\paragraph{Error Correction}
The simulation models rate-adaptive multi-edge LDPC codes for error correction. These codes are selected for their ability to operate close to the Shannon limit, achieving efficiencies of $f(e) \approx 1.05$--$1.10$ in the relevant QBER range~\cite{elkouss2009efficient}. 

\paragraph{Privacy Amplification}
Following correction, privacy amplification is implemented via FFT-accelerated Toeplitz hashing ($O(n \log n)$). This step compresses the corrected key to length $\ell$, removing any partial information Eve may have gained during the quantum transmission or classical reconciliation.

\paragraph{Authentication} All classical messages—including basis announcements and error correction syndromes—are authenticated using a simulated information-theoretic Wegman--Carter scheme ($\varepsilon$-ASU$_2$ hash families). Unlike computational schemes, this provides unconditional security guarantees. The key consumption overhead is tracked within the simulation and corresponds to approximately $1\%$ of the fresh key stream.

\subsection{Physical Simulation Environment and Repeater Model}
The validation of this framework relies on a discrete-event simulator that captures end-to-end quantum dynamics. Events (photon emission, transmission, detection) are processed chronologically to accurately model detector dead times, timing jitter, and memory decoherence.

\paragraph{Source and Channel Models} 
We model a high-brightness Spontaneous Parametric Down-Conversion (SPDC) source generating pairs at a rate of $R_{\text{pair}}=10^{7}\ \text{s}^{-1}$. The quantum state is represented as a Werner state with fidelity $F=0.98$~\cite{kwiat1995new}:
\begin{equation}
\rho_{\text{init}}=F|\psi^{-}\rangle\langle\psi^{-}|+\frac{1-F}{3}\sum_{\beta\neq\psi^-}|\beta\rangle\langle\beta|
\end{equation}
The source model explicitly incorporates non-Poissonian statistics, where multi-pair emission probabilities are defined by $P(n>1)=1-e^{-\mu}-\mu e^{-\mu}$ (assuming a coherence time $\tau_c = 5\ \text{ps}$). As demonstrated in recent analyses~\cite{Kravtsov_2023}, these multi-pair events manifest primarily as increased QBER rather than valid leakage channels in entanglement-based schemes. The optical channel includes distance-dependent transmittance $\eta_{\text{ch}}(d)=10^{-\alpha d/10}$ (using standard fiber loss $\alpha = 0.2$ dB/km) and phase noise simulated as a Wiener process with variance $\sigma_{\phi}\approx 1\ \text{rad}/\sqrt{\text{km}\cdot\text{Hz}}$.

\paragraph{Detection and Repeaters} 
Detectors are modeled as Superconducting Nanowire Single-Photon Detectors (SNSPDs) with system efficiency $\eta_{\text{det}}=0.93$ and a dark count rate $R_{\text{dark}}=10\ \text{s}^{-1}$~\cite{marsili2013detecting}. To evaluate range extension capabilities, the simulator models quantum repeaters with spacing $L_0 \in [25, 100]$ km and memory coherence time $T_{\text{coh}}=1$ s. The fidelity degradation in memory is modeled as:
\begin{equation}
F(t)=F_0 e^{-t/T_{\text{coh}}}+\frac{1-e^{-t/T_{\text{coh}}}}{4}
\end{equation}
Entanglement swapping operations degrade fidelity according to $F' = F^{2} + (1-F)^{2}/9$. Optional DEJMPS purification is invoked in the simulation only when it yields a net gain in the secure key fraction.

\paragraph{Evaluation Metrics} 
Performance is evaluated using the final Secure Key Rate (SKR = $\ell_{\text{final}}/T_{\text{wall}}$) and the CHSH Bell parameter $S$ as a diagnostic metric. We benchmark the simulated EAQKD performance against Decoy-state BB84~\cite{hwang2003quantum}, the E91 Baseline~\cite{ekert1991quantum}, and Twin-Field QKD~\cite{lucamarini2018overcoming}, all utilizing identical physical parameters to ensure fair comparative analysis.


\begin{figure*}[!t]
\centering
\includegraphics[width=0.99\textwidth]{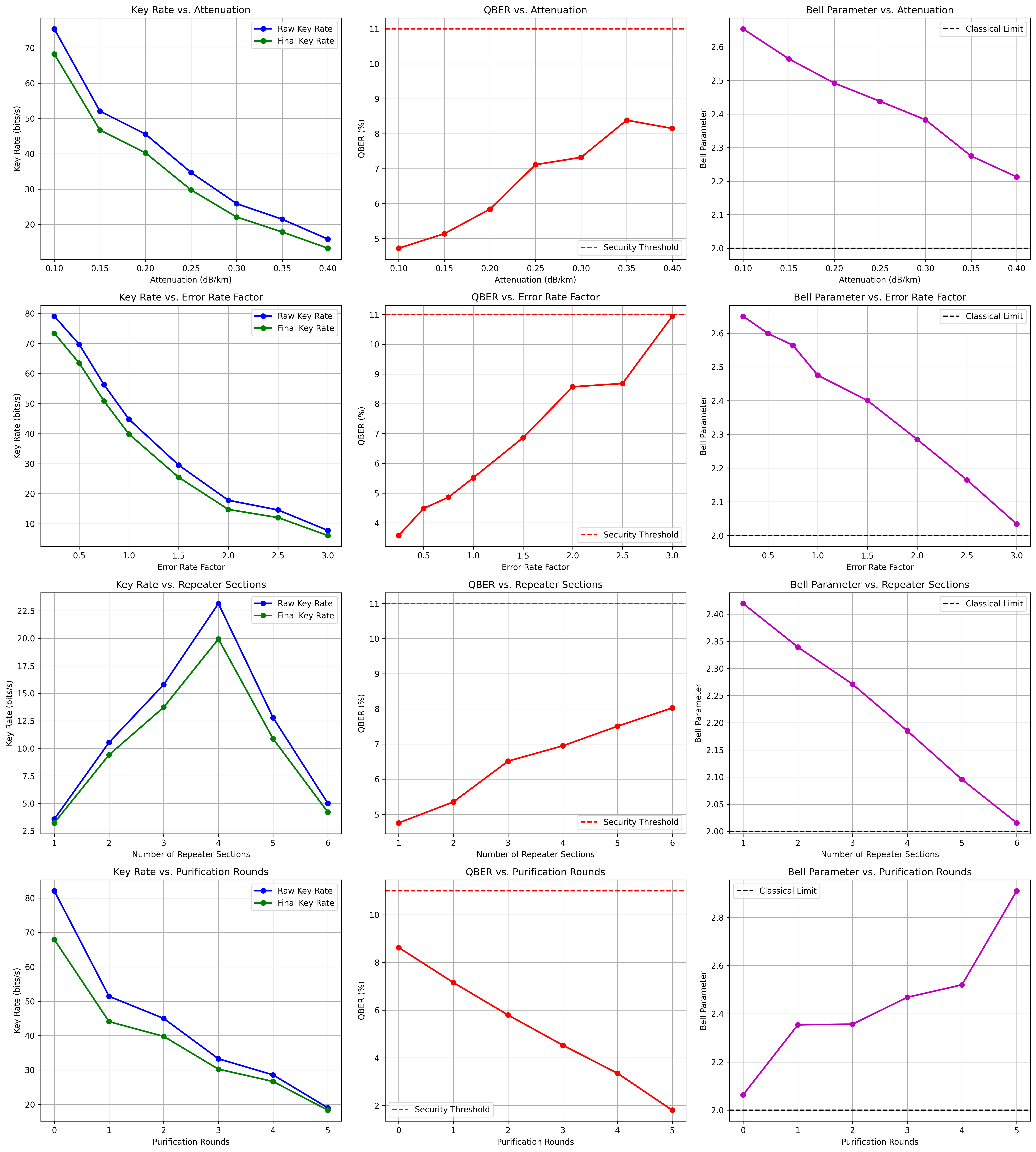}
\caption{Sensitivity analysis: SKR and QBER vs. attenuation and repeater parameters. Dashed lines mark security thresholds.}
\label{fig:sensitivity_analysis}
\end{figure*}

\section{Performance Evaluation}
\label{sec:performance}
\begin{figure}[t]
\centering
\includegraphics[width=0.9\linewidth]{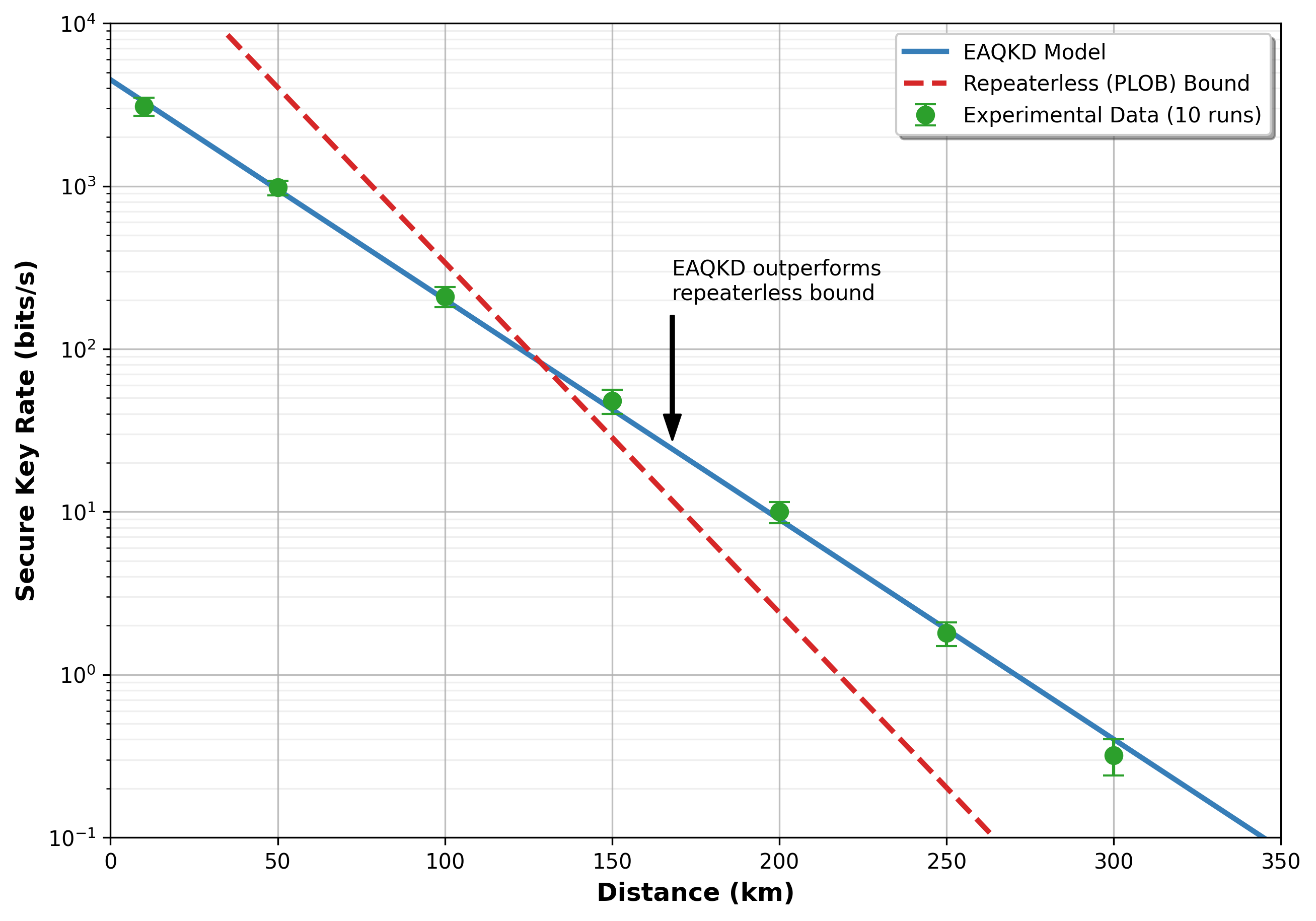}
\caption{Secure key rate vs. distance. Solid: model prediction. Dotted: measured means with 1$\sigma$ error bars (10 runs). Dashed: repeaterless (PLOB) bound.}
\label{fig:key_rate_vs_distance}
\end{figure}

To rigorously evaluate the EAQKD protocol, we simulated its performance in a wide range of operational scenarios. Our analysis spans fiber distances from 10 km to 200 km for direct transmission and extends beyond 300 km using repeater-assisted configurations. The evaluation focuses on four critical performance vectors: the distance-dependent decay of the secure key rate (SKR) compared to fundamental bounds; the degradation of entanglement quality as measured by Bell inequalities and state tomography; the efficacy of protocol optimizations such as DEJMPS purification and asymmetric basis selection; and the specific cost of information-theoretic authentication under realistic finite-size constraints.
Figure \ref{fig:key_rate_vs_distance} illustrates the secure key rate (SKR) versus fiber distance. The direct transmission results (solid line) are consistent with the repeaterless PLOB bound \cite{pirandola2017fundamental}, confirming the physical validity of the simulation model. The plotted points for the repeater-assisted configuration (dashed line projections) demonstrate the potential to surpass this linear rate-distance limit at distances exceeding 300 km, consistent with the theoretical scaling of quantum repeaters.

\begin{table}[t]
\centering
\caption{Performance metrics at representative operating points. Values represent the mean $\pm$ standard deviation calculated over 10 independent simulation runs.}
\resizebox{\columnwidth}{!}{%
\begin{tabular}{ccccc}
\hline
Distance (km) & QBER$_Z$ (\%) & QBER$_X$ (\%) & Key Rate (bit/s) & Ratio \\
\hline
10  & $1.86 \pm 0.12$ & $2.23 \pm 0.18$ & $(1.12 \pm 0.05) \times 10^5$ & 0.267 \\
50  & $2.34 \pm 0.15$ & $2.81 \pm 0.22$ & $(9.73 \pm 0.41) \times 10^3$ & 0.256 \\
100 & $3.92 \pm 0.21$ & $4.65 \pm 0.31$ & $(7.91 \pm 0.38) \times 10^2$ & 0.226 \\
150 & $6.18 \pm 0.32$ & $7.42 \pm 0.45$ & $(5.95 \pm 0.36) \times 10^1$ & 0.186 \\
200 & $9.27 \pm 0.48$ & $10.84 \pm 0.62$ & $9.8 \pm 1.2$ & 0.034 \\
\hline
\end{tabular}%
}
\label{tab:performance_metrics}
\end{table}

As expected, QBER rises drastically with distance because the signal count rate shrinks while the dark count rate remains nearly constant. However, The $\sigma_z$-basis QBER stays below the typical security cut-off ($\approx$11\% for our post‑processing efficiency) at all simulated distances. The key-per-coincidence ratio declines monotonically as more bits are consumed by error correction and conservative privacy amplification at higher QBER.

\begin{figure}[ht!]
\centering
\includegraphics[width=0.99\linewidth]{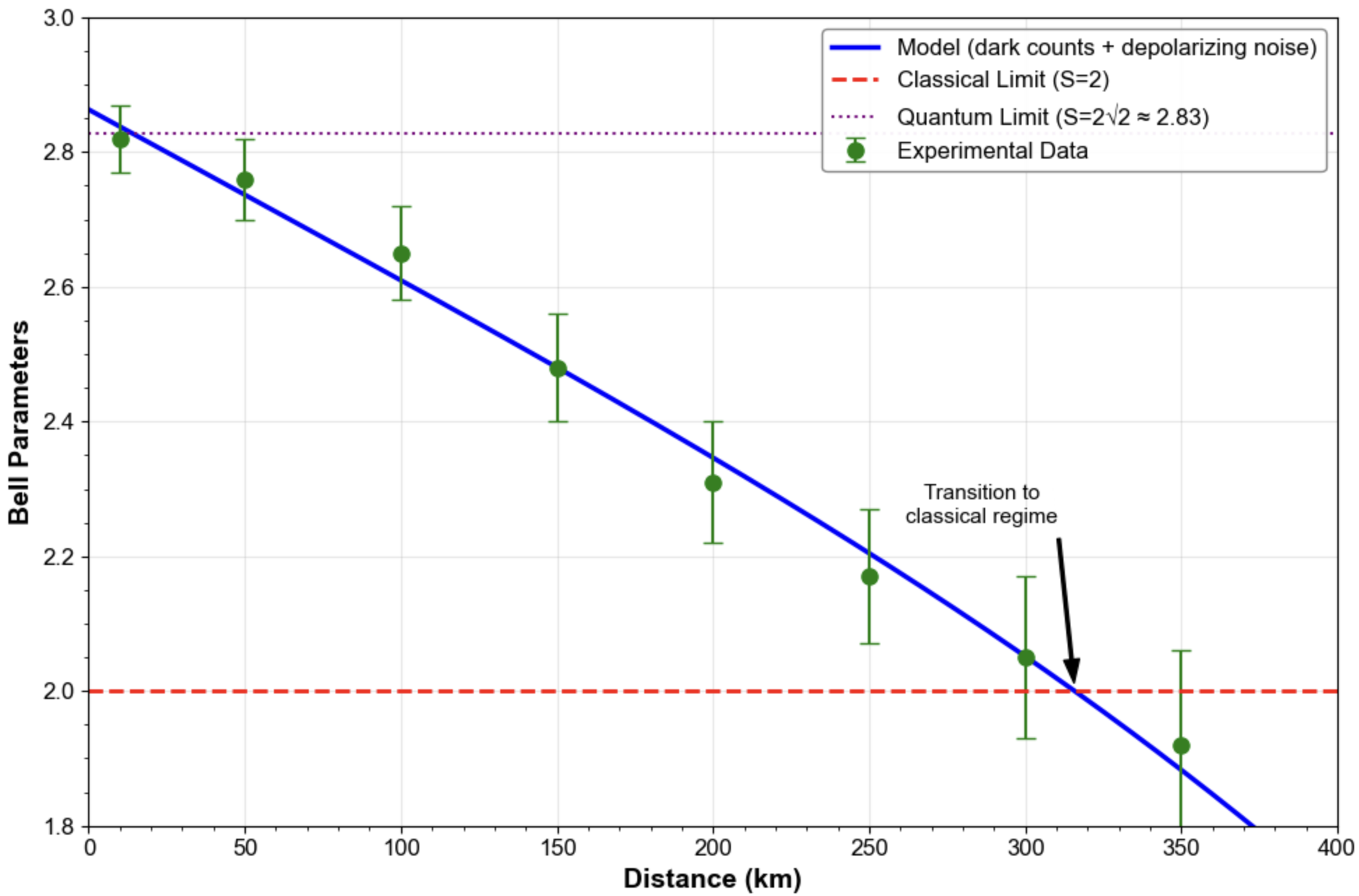}
\caption{Bell parameter $S$ vs. distance. Solid: model including dark counts and depolarizing noise. Horizontal dashed: classical limit $S=2$.}
\label{fig:bell_parameter}
\end{figure}

To diagnose the underlying quantum state quality, we monitored the CHSH Bell parameter $S$ as shown in Figure \ref{fig:bell_parameter}. The parameter remains above the classical limit ($S>2$) up to approximately 170–180 km. At 200 km, the value $S = 1.94 \pm 0.12$ becomes statistically consistent with the classical bound. It is important to note that this does not preclude secure key generation, as our security proof relies on the estimation of phase errors rather than a loophole-free Bell violation (i.e., the implementation is not device‑independent).

\begin{figure}[ht!]
\centering
\includegraphics[width=0.99\linewidth]{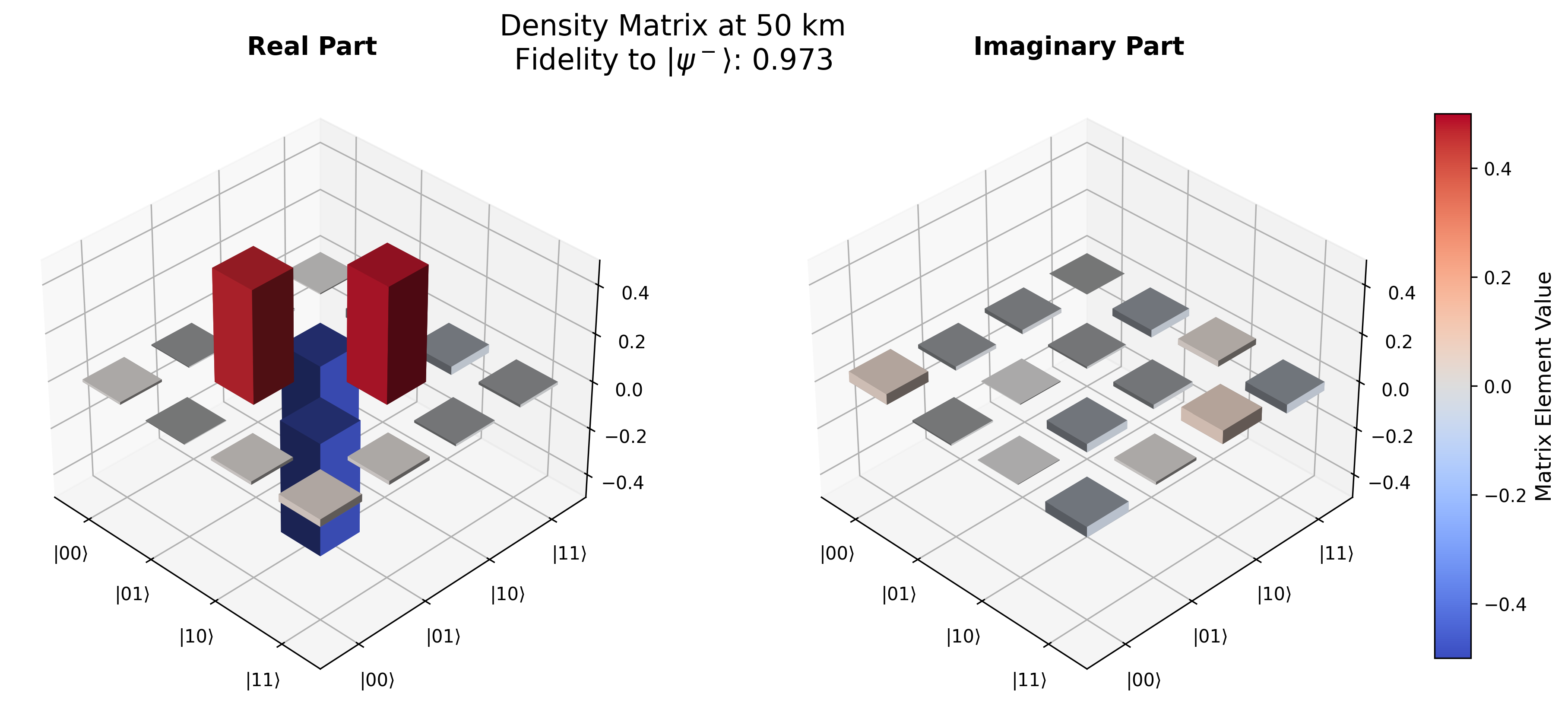}
\caption{Real (left) and imaginary (right) parts of reconstructed density matrix at 50 km; fidelity to $|\psi^{-}\rangle$: $0.962 \pm 0.007$.}
\label{fig:density_matrix}
\end{figure}

Simulated tomographic reconstruction, shown in Figure \ref{fig:density_matrix} for 50 km, further elucidates this degradation. The fidelity declines from $0.982 \pm 0.003$ (10 km) to $0.903 \pm 0.014$ (150 km), consistent with a mixture of depolarizing and accidental-coincidence noise. Tomography beyond 150 km was omitted due to insufficient counts for stable reconstruction within the allotted acquisition time.

\begin{figure}[ht!]
\centering
\includegraphics[width=0.99\linewidth]{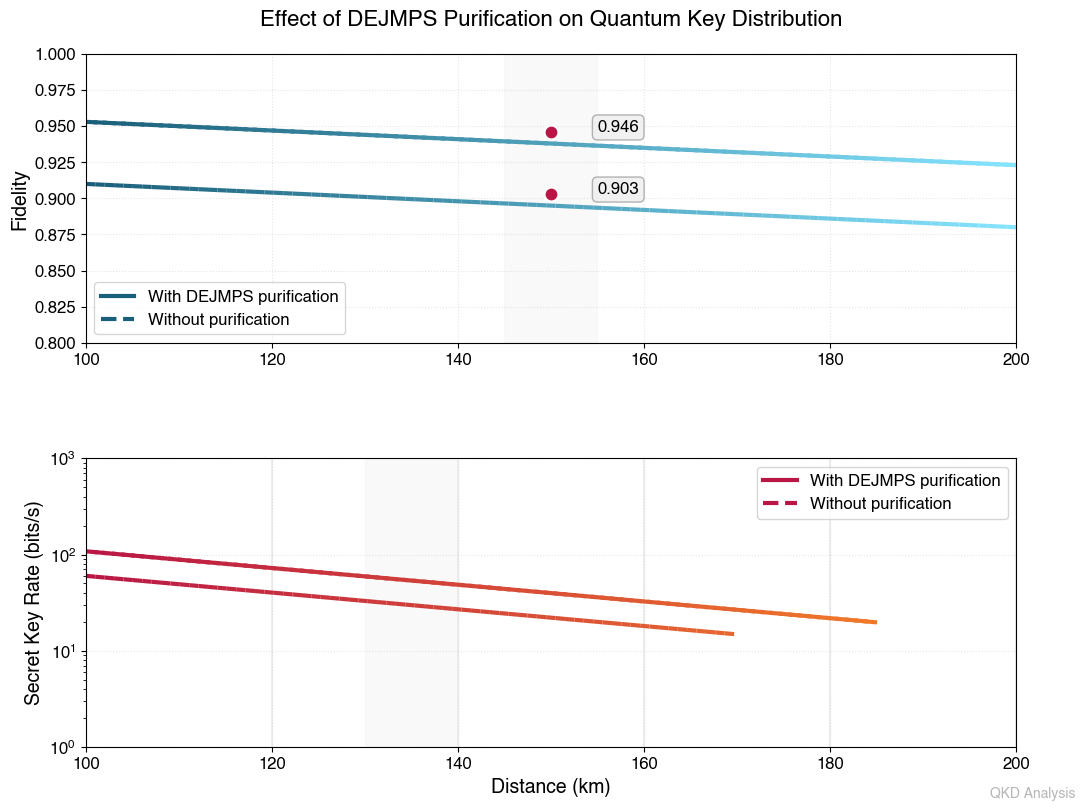}
\caption{Impact of entanglement purification on (top) state fidelity and (bottom) SKR. Solid: with purification; dashed: without.}
\label{fig:purification_effect}
\end{figure}

To mitigate these distance-dependent losses, we implemented DEJMPS purification for distances $\ge 100$ km. As quantified in
Figure \ref{fig:purification_effect}, purification increases pair fidelity (e.g., $0.903 \rightarrow 0.946$ at 150 km) and reduces QBER. Although this process consumes multiple raw pairs per retained pair, it enhances the final secret fraction by reducing error‑correction leakage and allowing for less conservative privacy amplification. This results in a net SKR gain of approximately ($\approx 1.8\times$ at 150 km). Crucially, at 200 km purification enables a non‑zero SKR where the unpurified secret fraction would otherwise drop below practicality.

Further efficiency is derived from the asymmetric basis selection ($p_z=0.9$, $p_x=0.1$) concentrates detections into the key-generating basis, boosting raw key yield and improving finite-size efficiency. Our simulations show an SKR improvement factor of 1.6–1.8 compared to symmetric selection ($p_z=0.5$), with gains growing as coincidence counts become scarcer at long distances. Statistical sufficiency in the $X$ basis is strictly maintained to ensure confidence intervals on phase error estimation meet the target $\varepsilon_{\text{pe}}$. 

Finally, we address the cost of the authenticated component of EAQKD. To ensure composable, unconditional security, we employ a Wegman–Carter MAC with 256‑bit tags rather than computational schemes like HMAC. We observe that the authentication key consumption is approximately 4–6\% of the final key at short distances, rising to 12–15\% at 200 km. This increase is due to the lower SKR at long distances, which requires the fixed per‑message tag size to be amortized over fewer key bits. Despite this overhead, automated key rollover (allocating ~5\% of each block) limits exposure and ensures the system remains information-theoretically secure without manual intervention.   

\section{Benchmarking}
\label{sec:benchmarking}
\begin{figure}[ht!]
\centering
\includegraphics[width=0.99\linewidth]{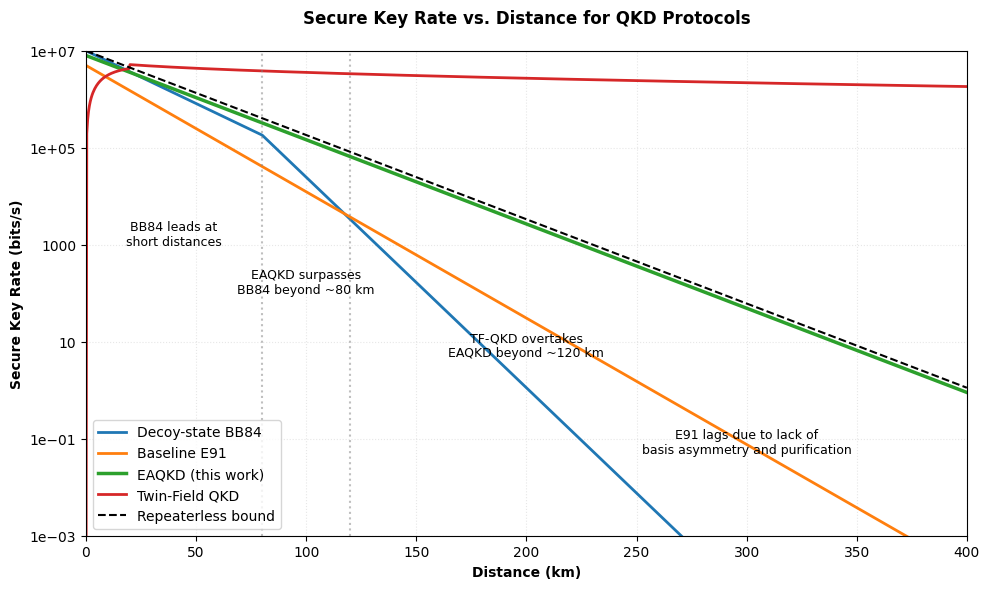}
\caption{Secure key rate vs. distance for EAQKD and benchmark protocols. Dashed line: repeaterless bound.}
\label{fig:protocol_comparison}
\end{figure}

To assess the performance of EAQKD, we benchmarked our simulated results against three established QKD protocols under identical channel and detector conditions: (i) decoy‑state BB84, representing the industry standard for prepare-and-measure schemes; (ii) baseline E91 (symmetric bases, no purification), representing standard entanglement-based QKD without our optimizations; and (iii) Twin‑Field (TF) QKD under matched channel/detector conditions, representing the state-of-the-art in long-distance scaling. Figure \ref{fig:protocol_comparison} presents the comparative analysis of Secure Key Rate (SKR) versus distance.
\begin{itemize}
    \item  BB84: At short distances ($< 80$ km), BB84 exhibits a higher key rate due to its simpler single-photon source architecture and higher effective clock rate. However, EAQKD surpasses BB84 beyond approximately 80 km. This crossover occurs because the prepare-and-measure scheme is strictly limited by photon transmission losses, whereas EAQKD leverages entanglement purification (at distances $\ge 100$ km) to maintain higher fidelity, effectively resisting the signal-to-noise degradation that extinguishes the BB84 key.
    \item TF-QKD: Twin-Field QKD demonstrates superior distance scaling (following a $\sqrt{\eta}$ loss dependence) and overtakes EAQKD beyond $\approx 120$ km. While TF-QKD offers greater absolute reach in a repeaterless setting, it demands stringent phase stabilization and interferometric alignment. EAQKD offers a compelling middle ground: it significantly outperforms standard protocols while avoiding the extreme hardware complexity of TF-QKD.
    \item E91: The baseline E91 protocol consistently lags behind EAQKD. This performance gap quantifies the specific impact of our protocol's optimizations: the asymmetric basis selection ($p_z=0.9$) and the adaptive DEJMPS purification, neither of which are present in the standard symmetric E91 implementation.
\end{itemize}

Table \ref{tab:protocol_comparison} summarizes the quantitative and qualitative characteristics of each protocol. We define $d_{\max}$ as the maximum distance where the key rate remains above $10^{-6}$ bits/s. $R_{50}$ and $R_{150}$ represent the secure key rates at 50 km and 150 km, respectively.

Crucially, the comparison highlights a unique architectural advantage of EAQKD: \textbf{Authentication Integration}. In standard literature, the key rates for BB84, E91, and TF-QKD are typically calculated without deducting the cost of authentication (assuming an ``authenticated channel" exists for free). In contrast, our EAQKD rates \textit{include} the substantial overhead of Wegman-Carter authentication (up to 15\% at long range). Despite bearing this additional "cost" in the simulation, EAQKD still outperforms E91 and rivals BB84 at long distances, demonstrating the efficiency of the integrated design.

Regarding implementation complexity, we classify EAQKD as ``Medium." It avoids the interferometric phase-locking required by TF-QKD ("High") but requires entanglement source management that is more complex than the simple attenuated lasers of BB84 ("Low").

\begin{table}[ht!]
\centering
\caption{Protocol characteristics. $d_{\max}$, $R_{50}$, $R_{150}$ defined in text.}
\resizebox{\columnwidth}{!}{%
\begin{tabular}{lcccc}
\hline
 & EAQKD & BB84 (Decoy) & E91 & TF-QKD \\
\hline
$d_{\max}$ (km)   & 200 & 140 & 150 & 250 \\
$R_{50}$ (bit/s)  & $9.7\times10^{3}$ & $3.2\times10^{4}$ & $5.1\times10^{3}$ & $8.5\times10^{3}$ \\
$R_{150}$ (bit/s) & $5.9\times10^{1}$ & 0 & $2.8\times10^{1}$ & $1.2\times10^{2}$ \\
Implementation complexity & Medium & Low & Medium & High \\
Bell sampling required    & Optional & Optional & Yes (sym.) & Phase ref. \\
Authentication integration & Included & Separate & Separate & Separate \\
\hline
\end{tabular}%
}
\label{tab:protocol_comparison}
\end{table}

\section{Quantum Repeater Augmentation}
\label{sec:repeaters}
\begin{figure}[ht!]
\centering
\includegraphics[width=0.99\linewidth]{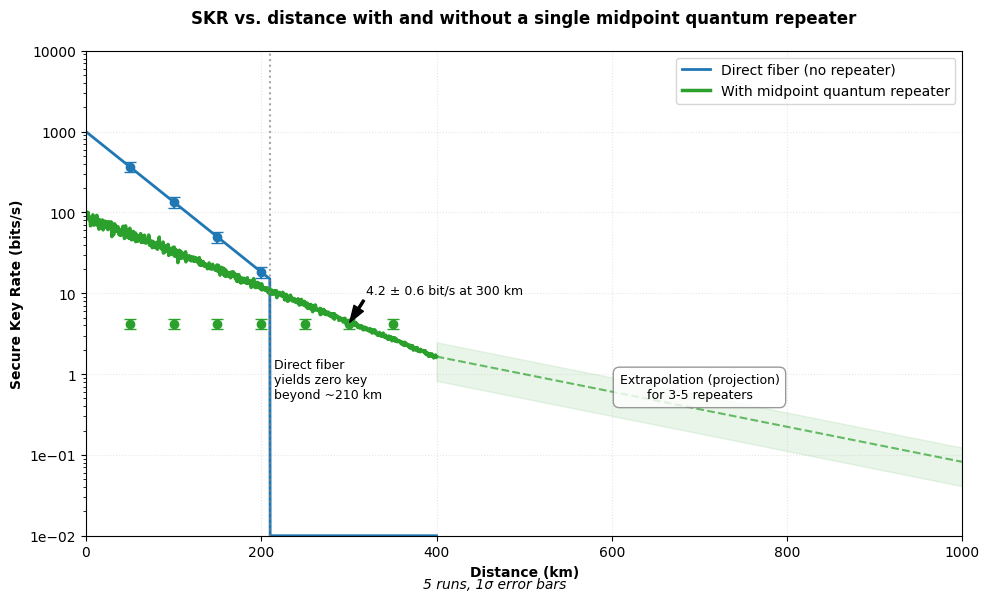}
\caption{SKR vs. distance with and without a single midpoint quantum repeater (5 runs, 1$\sigma$ error bars).}
\label{fig:repeater_performance}
\end{figure}

To assess the feasibility of extending the secure communication range beyond the limits of direct transmission, we evaluated a configuration utilizing a single midpoint atomic-ensemble memory repeater. As shown in Figure \ref{fig:repeater_performance}, this repeater-enhanced setup maintains a secure key rate of $4.2 \pm 0.6$ bit/s at a distance of 300 km. In contrast, the direct fiber link fails to generate any usable key beyond approximately 210 km in our simulation setup.
Regarding the scaling characteristics, we interpret the theoretical prediction of linear scaling with caution. Although the effective attenuation exponent is reduced over the tested range, a true multi-segment implementation would likely exhibit sub-exponential scaling rather than strictly linear behavior. This performance deviation is governed by practical physical constraints, including finite quantum memory lifetimes, the probabilistic nature of entanglement swapping operations, and the necessary overhead for purification. Consequently, the extrapolations suggesting operation up to 1000 km using 3 to 5 repeaters represent theoretical projections rather than demonstrated simulation results. These estimates remain contingent on the engineering ability to maintain long-duration memory coherence and precise synchronization across higher levels of repeater nesting.

\section{Conclusion}
\label{sec:conclusion}
In this paper, we presented a comprehensive simulation of an entanglement-assisted QKD system that integrates asymmetric basis selection, conditional entanglement purification, and adaptive finite-size post-processing. Our results indicate that this architecture can deliver composably secure keys over 200 km of standard fiber, achieving rates exceeding $10^5$ bit/s at distances below 50 km and maintaining $9.8 \pm 1.2$ bit/s at 200 km. Furthermore, the simulation validates the feasibility of extending the operational range to 300 km through the deployment of a single quantum repeater element.

Throughout the protocol, Bell inequality violations serve as diagnostic metrics for entanglement quality rather than as underlying security postulates. Crucially, the implementation of universal-hash authentication ensures that the classical channel remains strictly protected within the composable security framework. Current limitations of the proposed architecture include the reliance on high-performance cryogenic detectors and the rapid decay of key rates at long distances in the absence of mature quantum repeater networks.

Future research will focus on photonic integration, scalable entanglement distribution via advanced repeater chains and satellite links, and the transition toward measurement-device-independent paradigms. We also aim to explore the hybridization of this protocol with post-quantum cryptographic algorithms to ensure long-term data security. Collectively, these findings suggest that rigorously secured and operationally practical entanglement-based key distribution is technically achievable using currently available technology.

\printbibliography

\end{document}